\documentclass[xcolor=pdftex,usenames,dvipsnames,table]{PoS}

\usepackage{slashed}
\usepackage{graphicx}
\usepackage{amssymb,amsmath} \usepackage{pifont}
\usepackage{multirow,lscape}
\usepackage{array}
\usepackage[lofdepth,lotdepth]{subfig}
\usepackage{verbatim}
\usepackage{bm}
\usepackage{comment}

\graphicspath{{./figs/}}

\providecommand{\abs}[1]{\lvert#1\rvert}

\providecommand{\wbar}[1]{\overline#1}

\providecommand{\ket}[1]{\lvert#1\rangle}

\providecommand{\epsK}{\varepsilon_K}

\newcolumntype{C}[1]{>{\centering\arraybackslash}p{#1}}

\definecolor{dkgray}{RGB}{145,145,145}
\definecolor{violet}{RGB}{50,0,200}

\title{%
Current status of $\epsK$ calculated with lattice QCD inputs
}

\ShortTitle{%
Current status of $\epsK$
}

\author{%
  Jon A.~Bailey, Yong-Chull Jang, and \speaker{Weonjong Lee} \\
  Lattice Gauge Theory Research Center, CTP, and FPRD, \\
  Department of Physics and Astronomy, \\ 
  Seoul National University,
  Seoul, 151-747, South Korea\\
  E-mail: \email{wlee@snu.ac.kr}
}

\author{SWME Collaboration}

\abstract{ We present results for $\epsK$, the indirect CP violation
  parameter, calculated in the Standard Model using inputs from
  lattice QCD: the kaon bag parameter $\hat{B}_K$, and the CKM matrix
  element $V_{cb}$ from the axial current form factor for the
  exclusive decay $\bar{B}\to D^*\ell\bar{\nu}$ at zero-recoil.
  In addition, we take the coordinates of the unitarity triangle apex
  $(\bar{\rho},\bar{\eta})$ from the angle-only fit of the UTfit
  Collaboration and use $V_{us}$ to fix $\lambda$.
  In order to estimate the systematic error, we also use Wolfenstein
  parameters from the CKMfitter and UTfit.
  We find a $3.3(2)\sigma$ difference between $\epsK$ and experiment
  with exclusive $V_{cb}$.
  We report details of this preliminary result.
}

\FullConference{%
  The 32nd International Symposium on Lattice Field Theory\\
  23-28 June, 2014\\
  Columbia University, New York, NY}

\begin{document}

\section{Introduction}
Indirect CP violation in the neutral kaon system is parametrized by
$\epsK$
\begin{equation}
  \epsK 
  \equiv \frac {A[K_L \to \pi\pi(I=0)]} {A[K_S \to \pi\pi(I=0)]} \,.
\end{equation}
%
%
%
%
%
Experimentally \cite{Beringer2012:PhysRevD86.010001},
\begin{equation}
  \label{eq:epsK-exp}
  \epsK = (2.228 \pm 0.011) \times 10^{-3} 
  \times e^{i\phi_\epsilon}\,, \qquad
  \phi_\epsilon = 43.52 \pm 0.05 {}^\circ\,.
\end{equation}

We can also calculate $\epsK$ in the Standard Model (SM).
In the SM, the CP violation comes solely from a single phase in the
CKM matrix elements \cite{ Cabibbo1963:PhysRevLett.10.531,
  Kobayashi1973:ProgTheorPhys.49.652}.
The SM allows the mixing of neutral kaons $K^0$ and $\wbar{K}^0$
through loop processes, and describes contributions to the mass
splitting $\Delta M_K$ and $\epsK$.
Hence, we can test the SM through the CP violation by comparing
the experimental and theoretical values of $\epsK$.

We can express $\epsK$ in terms of input parameters from lattice QCD
and experiments.
Among them, the input parameters $\hat{B}_{K}$ and $V_{cb}$ long
dominated the statistical and systematic uncertainty in the SM
evaluation of $\epsK$.
During the past decade, lattice QCD has reduced the $\hat{B}_K$ error
dramatically, to $\approx 1.3\%$.
The average of the lattice results is available from Flavour Lattice
Averaging Group (FLAG)~\cite{Aoki2013:hep-lat.1310.8555}.
We calculate $\epsK$ using the lattice average for $\hat{B}_K$ from
FLAG and compare the value of $\epsK$ calculated with the updated
result for $\hat{B}_K$ from the SWME Collaboration, which has a larger
uncertainty of $\approx 5\%$~\cite{Bae2014:prd.89.074504}.

There exists a $3\sigma$ difference in $V_{cb}$ between exclusive
and inclusive channels \cite{GS}.
Our analysis shows how this discrepancy propagates to $\epsK$.
The axial current form factor for the semi-leptonic decay $\bar{B}\to
D^*\ell\bar{\nu}$ at zero recoil, with the experimental branching
fraction, can be used to determine $V_{cb}$.  The Fermilab Lattice and
MILC Collaborations (FNAL/MILC) have updated their lattice calculation
of the form factor~\cite{JWL}.
We compare $\epsK$ obtained using the exclusive $V_{cb}$ from the
FNAL/MILC result with $\epsK$ obtained using the inclusive $V_{cb}$ in
Ref.~\cite{GS}.

We use the Wolfenstein parametrization for the CKM matrix, truncating
the series at $\mathcal{O}(\lambda^7) \approx 10^{-5}$.
We examine three different choices of Wolfenstein parameters: (1)
$\lambda$, $\bar{\rho}$, and $\bar{\eta}$ from the global unitarity
triangle (UT) fit of CKMfitter, (2) $\lambda$, $\bar{\rho}$, and
$\bar{\eta}$ from the global UT fit of UTfit, and (3) $\bar{\rho}$ and
$\bar{\eta}$ from an angle-only UT fit from UTfit, with $\lambda$ from
$V_{us}$ \cite{Beringer2012:PhysRevD86.010001,Bevan2013:npps241.89}.
In all cases we take $V_{cb}$ instead of $A$.
The angle-only fit (AOF) does not use $\epsK$, $\hat{B}_K$, and
$V_{cb}$ to determine the UT apex $\bar{\rho}$ and $\bar{\eta}$.
Hence, it provides a way to test the validity of the SM with $\epsK$,
using the lattice results of $\hat{B}_K$ and $V_{cb}$.

To estimate the effect of correlations in lattice input parameters, we
note that the $V_{cb}$ dominates the error in $\epsK$, and
the FLAG $\hat{B}_K$ is dominated by the BMW result
\cite{Durr:2011ap}.
The correlation between the BMW $\hat{B}_K$ and the exclusive $V_{cb}$
from the FNAL/MILC form factor is negligible.
Hence, we assume that the correlation between the lattice input
parameters $\hat{B}_K$, $V_{cb}$ and $\xi_0$ are negligible.
To determine the value of $\epsK$, we take uncorrelated inputs for all
the parameters, and use the Monte Carlo method to determine the error.
We also compare the results with standard error propagation to
cross-check them.
In the error budget, we quote results obtained using the error
propagation method.

\section{Indirect CP Violation in the Kaon System: $\epsK$}

We use the master formula in Eq.~\eqref{eq:epsK_SM_0} to evaluate the
SM value of $\epsK$.
\begin{align}
  \epsK^{SM}
  &= \frac{\tilde{\varepsilon} + i \xi_0}{1 + i \tilde{\varepsilon} \xi_0} 
  = \tilde{\varepsilon}_0 + i\xi_0 + \mathcal{O}(\tilde{\varepsilon}^3_0) 
  \label{eq:epsK_0} \\
  &= e^{i\theta} \sqrt{2}\sin{\theta} 
     \Big( C_{\epsilon} \hat{B}_{K} X + \xi_{0} \Big) +
     \xi_\text{LD} + \mathcal{O}(\tilde{\varepsilon}^3_0)
  \label{eq:epsK_SM_0}
\end{align}
where $\tilde\varepsilon = \tilde\varepsilon_0(1 +
\tilde\varepsilon^2)$ \cite{Jang2012:PoS.LAT2012.269,Wlee}, $\xi_0$ is
defined in Eq.~\eqref{eq:xi0-from-epsprime}, and $\xi_\text{LD}$ is
the long distance effect of $\approx 2\%$, which we neglect in
this paper.
We also neglect the truncation error of
$\mathcal{O}(\tilde{\varepsilon}_0^3) \cong 10^{-9}$.
The mixing parameter $\tilde{\varepsilon}$ is defined by the following.
\begin{align}
  \ket{K_{S}} =& \frac{1}{ \sqrt{ 1+\abs{\tilde{\varepsilon}}^2 } }
  (\ket{K_{1}} + \tilde{\varepsilon} \ket{K_{2}}) 
  \,,\qquad
  \ket{K_{L}} = \frac{1}{ \sqrt{ 1+\abs{\tilde{\varepsilon}}^2 } }
  (\ket{K_{2}} + \tilde{\varepsilon} \ket{K_{1}}) \,,
\end{align}
where $\ket{K_{1}}$ and $\ket{K_{2}}$ are CP even and odd states,
respectively.
In our phase convention of $CP \ket{K^{0}} = - \ket{\wbar{K^{0}}}\,$,
they are
\begin{equation}\label{eq:def of K1K2}
  \vert K_{1} \rangle
  = \frac{1}{\sqrt{2}}
  \big( \ket{K^{0}} - \ket{\wbar{K^{0}}} \big) \,,
  \qquad \qquad
  \vert K_{2} \rangle
  = \frac{1}{\sqrt{2}}
  \big( \ket{K^{0}} + \ket{\wbar{K^{0}}} \big) \,.
\end{equation}
The factor $X$ is 
\begin{align}\label{eq:SD-box-kaon}
  X &= \bar{\eta}\lambda^{2} \abs{V_{cb}}^{2}
     \Big[ \abs{V_{cb}}^{2} (1-\bar{\rho})
     \eta_{2} S_{0}(x_{t})     
   + \eta_{3} S_{0}(x_{c},x_{t}) - \eta_{1} S_{0}(x_{c}) \Big]
\end{align}
where $x_i = m_i^2/M_W^2$ with $(i = c, t)$, and $S_0$'s are the
Inami-Lim functions.
$X$ takes into account the short-distance contribution
of the box-diagram \cite{Buras1998:hep-ph/9806471}.
We use the experimental value for $\Delta M_K$ because the theoretical
value does not have enough precision yet \cite{RBC2014:PRL}.
Other input parameters which appear in Eq.~\eqref{eq:SD-box-kaon}, the
factor $C_{\varepsilon}$, 
\begin{gather}
  C_{\varepsilon}
  = \frac{ G_{F}^{2} F_K^{2} m_{K^{0}} M_{W}^{2} }
       { 6\sqrt{2} \pi^{2} \Delta M_{K} } \,,
\end{gather}
and $\hat{B}_K$ will be explained in the next section.

\section{Input Parameters}
\label{sec:iparam}

The input values that we use for $V_{cb}$ are summarized in
Table~\ref{tbl:in-Vcb}. The inclusive determination considers the
following inclusive decays: $ B \to X_c l \nu \,,$ and $B \to X_s
\gamma $.
Moments of lepton energy, hadron masses, and photon energy are
measured from the relevant decay.
Those moments are fit to the theoretical expressions which are
obtained by applying the operator product expansion (OPE) to the decay
amplitude with respect to the strong coupling $\alpha_s$, and inverse
heavy quark mass $\Lambda/m_b$.
There are two schemes for the choice of $b$ quark mass $m_b$ in the
heavy quark expansion: kinetic scheme and 1S scheme.
We use the value obtained using the kinetic scheme, which has somewhat
larger errors \cite{Beringer2012:PhysRevD86.010001}.

The exclusive determination considers the semi-leptonic decay of
$\bar{B}$ to $D$ or $D^{\ast}$.
Here, we use the most up-to-date value from FNAL/MILC lattice
calculation of the form factor of the semi-leptonic decay $\bar{B}\to
D^{\ast}\ell\bar{\nu}$ at zero-recoil \cite{JWL}.

Several lattice calculations of $\hat{B}_K$ are available.
FLAG summarizes lattice results with $N_f=2+1$ and provides the
lattice average.
Here, we use the $N_f=2+1$ FLAG average~\cite{
  Bae2012:PhysRevLett.109.041601, rbc-prd-2011-1, alv-prd-2010-1,
  Durr:2011ap, Aoki2013:hep-lat.1310.8555} and SWME calculation as
inputs, Table~\ref{tbl:in-BK}.
FLAG uses the previous $\hat{B}_K$ result of SWME collaboration
\cite{Bae2012:PhysRevLett.109.041601}, and it is not much different
from the updated value \cite{Bae2014:prd.89.074504} that we use in
this analysis.

\begin{table}[t!]\footnotesize
  \centering
  \renewcommand{\arraystretch}{1.2}
  \subfloat[\label{tbl:in-Vcb}]{
  \begin{tabular}{C{1cm}|p{2.5cm}|p{1.4cm}}
  \hline\hline
  \multirow{2}{*}{$\abs{V_{cb}}$} 
  & $42.42(86)$
  &\cite{GS} Incl. \\ \cline{2-3}
  & $39.04(49)(53)(19)$
  &\cite{JWL} Excl. \\
  \hline\hline
  \end{tabular}
  }\hspace{0.5cm}
  \subfloat[\label{tbl:in-BK}]{
  \begin{tabular}{C{1cm}|p{2.5cm}|p{1.7cm}}
	\hline\hline
	\multirow{2}{*}{$\hat{B}_{K}$} 
	& $0.7661(99)$
	&\cite{Aoki2013:hep-lat.1310.8555} FLAG \\ \cline{2-3}
	& $0.7379(47)(365)$
	&\cite{Bae2014:prd.89.074504} SWME \\
	\hline\hline
	\end{tabular}
  }
  \caption{ The magnitudes of inclusive and exclusive $V_{cb}$ are
    given in units of $10^{-3}$. The inclusive $V_{cb}$ value is
    determined in the kinetic scheme for the heavy quark expansion.}
  \label{tab:Vcb+BK}
\end{table}

The CKMfitter and UTfit groups provide the Wolfenstein parameters
$\lambda, \bar{\rho}, \bar{\eta}$ and $A$ from the global UT fit.
Here, we use $\lambda, \bar{\rho}, \bar{\eta}$ from CKMfitter and
UTfit, and we use $V_{cb}$ instead of $A$ when we calculate $\epsK$
as in Eq.~\eqref{eq:SD-box-kaon}.
\begin{equation}
  \abs{V_{cb}} = A \lambda^2 + \mathcal{O}{(\lambda^7)}\,,
\end{equation}
where $\mathcal{O}(\lambda^7)\approx 2\times 10^{-5}$ is negligible.
The parameters $\lambda$, $\bar{\rho}$, and $\bar{\eta}$ are collected
in the Table~\ref{tbl:in-wolf}.

The parameters $\epsilon_K, \hat{B}_K$, and $V_{cb}$ are inputs to the
global UT fit.
Hence, the Wolfenstein parameters extracted from the global UT fit of
the CKMfitter and UTfit groups contain unwanted dependence on the
$\epsilon_K$ calculated from the master formula,
Eq.~\eqref{eq:epsK_SM_0}.
To self-consistently determine $\epsK$, we take another input set from
the angle-only fit (AOF).
The AOF does not use $\epsilon_K, \hat{B}_K$, and $V_{cb}$ as inputs
to determine the UT apex of $\bar{\rho}$ and $\bar{\eta}$
\cite{Bevan2013:npps241.89}.
The AOF gives the UT apex $(\bar{\rho}, \bar{\eta})$ but not
$\lambda$.
We can take $\lambda$ independently from the CKM matrix element
$V_{us}$, because this is parametrized by
\begin{equation}
  \abs{V_{us}} = \lambda + \mathcal{O}(\lambda^7) \,.
\end{equation}
Here we use the average of results extracted from the $K_{\ell3}$ and
$K_{\mu2}$ decays \cite{Beringer2012:PhysRevD86.010001}.  

The RBC-UKQCD collaboration provides lattice results of
$\mathrm{Im}\ A_2$ and $\xi_0$ \cite{Blum2011:PhysRevLett.108.141601}.
They obtain $\xi_0$ using the relation
\begin{equation}
  \label{eq:xi0-from-epsprime}
  \mathrm{Re} \Big( \frac{\epsilon^\prime_{K}}{\epsilon_{K}} \Big) 
  =	\frac{\cos(\phi_{\epsilon^\prime}-\phi_\epsilon)}
         {\sqrt{2}\abs{\epsilon_{K}}}  
		\frac{\mathrm{Re}\ A_{2}}{\mathrm{Re}\ A_{0}} 
		\Big( \frac{\mathrm{Im}\ A_{2}}{\mathrm{Re}\ A_{2}} 
                - \xi_{0} \Big) \,,\qquad
  \xi_0 \equiv \frac{\mathrm{Im}\ A_{0}}{\mathrm{Re}\ A_{0}} \,.
\end{equation}
In using this relation, input parameters except $\xi_0$ and
$\mathrm{Im}\ A_2$ are taken from the experimental values, as
suggested in Ref.~\cite{Blum2011:PhysRevLett.108.141601}.
In particular, they use the experimental value of $\epsK$ as an input
parameter to determine $\xi_0$.
However, the error is dominated by the experimental error of
$\mathrm{Re}(\epsilon^\prime_{K}/\epsK)$ $\approx 14\%$.
In the numerator, $\cos(\phi_{\varepsilon^\prime} - \phi_\varepsilon)$
is approximated by $1$, because the two phases are very close to each
other.
The result of $\xi_0$ is given in Table~\ref{tbl:in-xi0-others}.

The remaining input parameters are the Fermi constant $G_F$, $W$ boson
mass $M_W$, quark masses $m_q$, kaon mass $m_K^0$, mass difference
$\Delta M_K$, kaon decay constant $F_K$, and QCD short distance
correction factors $\eta_i$; these are summarized in
Table~\ref{tbl:in-xi0-others}.
The factors $\eta_1$ and $\eta_2$ are next-to-leading order (NLO)
results.\footnote{The NNLO result of $\eta_1$ ($=\eta_{cc}$) is
  available in Ref.~\cite{Brod:2011ty}. However, there is a claim that
  the error is overestimated \cite{Buras:2013raa}. This issue is under
  further investigation. We plan to address this issue in
  Ref.~\cite{Wlee}.}
Recently, the next-to-next-to-leading order (NNLO) calculation became
available for $\eta_3$~\cite{Brod2010:prd.82.094026}, and we take this
value as an input.
\begin{table}[t!]\footnotesize
  \centering
  \renewcommand{\arraystretch}{1.2}
  \subfloat[Wolfenstein Parameters\label{tbl:in-wolf}]{
  \begin{tabular}{C{0.5cm}|p{2cm}|p{2.7cm}}
  \hline\hline
  \multirow{3}{*}{$\lambda$} 
  & $0.22535(65)$ 
  &\cite{Beringer2012:PhysRevD86.010001} CKMfitter \\ \cline{2-3}
  & $0.22535(65)$ 
  &\cite{Beringer2012:PhysRevD86.010001} UTfit \\ \cline{2-3}
  & $0.2252(9)$
  &\cite{Beringer2012:PhysRevD86.010001} $\abs{V_{us}}$ (AOF) \\ \cline{1-3}
  \multirow{3}{*}{$\bar{\rho}$}
  & $\displaystyle 0.131^{+0.026}_{-0.013}$ 
  &\cite{Beringer2012:PhysRevD86.010001} CKMfitter \\ \cline{2-3}
  & $0.136(18)$ 
  &\cite{Beringer2012:PhysRevD86.010001} UTfit \\ \cline{2-3}
  & $0.130(27)$
  &\cite{Bevan2013:npps241.89} UTfit (AOF) \\ \cline{1-3}
  \multirow{3}{*}{$\bar{\eta}$} 
  & $\displaystyle 0.345^{+0.013}_{-0.014}$ 
  &\cite{Beringer2012:PhysRevD86.010001} CKMfitter \\ \cline{2-3}
  & $0.348(14)$ 
  &\cite{Beringer2012:PhysRevD86.010001} UTfit\\ \cline{2-3}
  & $0.338(16)$
  &\cite{Bevan2013:npps241.89} UTfit (AOF) \\ \cline{1-3}
  \hline\hline
  \end{tabular}
  }\hspace{0.5cm}
  \subfloat[\label{tbl:in-xi0-others}]{
  \begin{tabular}{C{1cm}|p{4.5cm}|p{0.5cm}}
	\hline\hline
	$G_{F}$ 
	& $1.1663787(6) \times 10^{-5}$ GeV$^{-2}$ 
	&\cite{Beringer2012:PhysRevD86.010001} \\ \cline{1-3}
	$M_{W}$ 
	& $80.385(15)$ GeV 
	&\cite{Beringer2012:PhysRevD86.010001} \\ \cline{1-3}
	$m_{c}(m_{c})$ 
	& $1.275(25)$ GeV 
	&\cite{Beringer2012:PhysRevD86.010001} \\ \cline{1-3}
	$m_{t}(m_{t})$ 
	& $163.3(2.7)$ GeV 
	&\cite{Alekhin2012:plb.716.214} \\ \cline{1-3}
	$\eta_{1}$ 
	& $1.43(23)$ 
	&\cite{Buras2008:PhysRevD.78.033005} \\ \cline{1-3}
	$\eta_{2}$ 
	& $0.5765(65)$ 
	&\cite{Buras2008:PhysRevD.78.033005} \\ \cline{1-3}
	$\eta_{3}$ 
	& $0.496(47)$ 
	&\cite{Brod2010:prd.82.094026} \\ \cline{1-3}
	$\theta$ 
	& $43.52(5)^{\circ}$ 
	&\cite{Beringer2012:PhysRevD86.010001} \\ \cline{1-3}
	$m_{K^{0}}$ 
	& $497.614(24)$ MeV 
	&\cite{Beringer2012:PhysRevD86.010001} \\ \cline{1-3}
	$\Delta M_{K}$ 
	& $3.484(6) \times 10^{-12}$ MeV 
	&\cite{Beringer2012:PhysRevD86.010001} \\ \cline{1-3}
	$F_K$
	& $156.1(8)$ MeV 
  &\cite{Beringer2012:PhysRevD86.010001} \\ \cline{1-3}
	$\xi_{0}$ 
	& $-1.63(19)(20) \times 10^{-4}$ 
	&\cite{Blum2011:PhysRevLett.108.141601} \\
	\hline\hline
  \end{tabular}
  }
  %
  \caption{ Wolfenstein parameters, $\xi_0$, and other inputs. }
  \label{tbl:in-Wolf-xi0-others}
\end{table}

\section{Results} 
\label{sec:results}
We use the Monte Carlo method to calculate the value of $\epsilon_K$
in the SM.
Assuming the input parameters are uncorrelated with each other and
follow the Gaussian distribution with mean and standard deviation
given in Tables \ref{tab:Vcb+BK} and \ref{tbl:in-Wolf-xi0-others}, we
generate $10^5$ random sample vectors.
The dimension of a sample vector is $n=17$, the total number of input
parameters which appear in the $\epsK$ master formula of
Eq.~\eqref{eq:epsK_SM_0}.

\begin{figure*}[t!]
  \centering
  \vspace*{-5mm}
  \subfloat[Incl. $V_{cb}$ + FLAG $\hat{B}_K$\label{sfig:incl.avg.AO}]{%
    \includegraphics[width=0.4\textwidth]{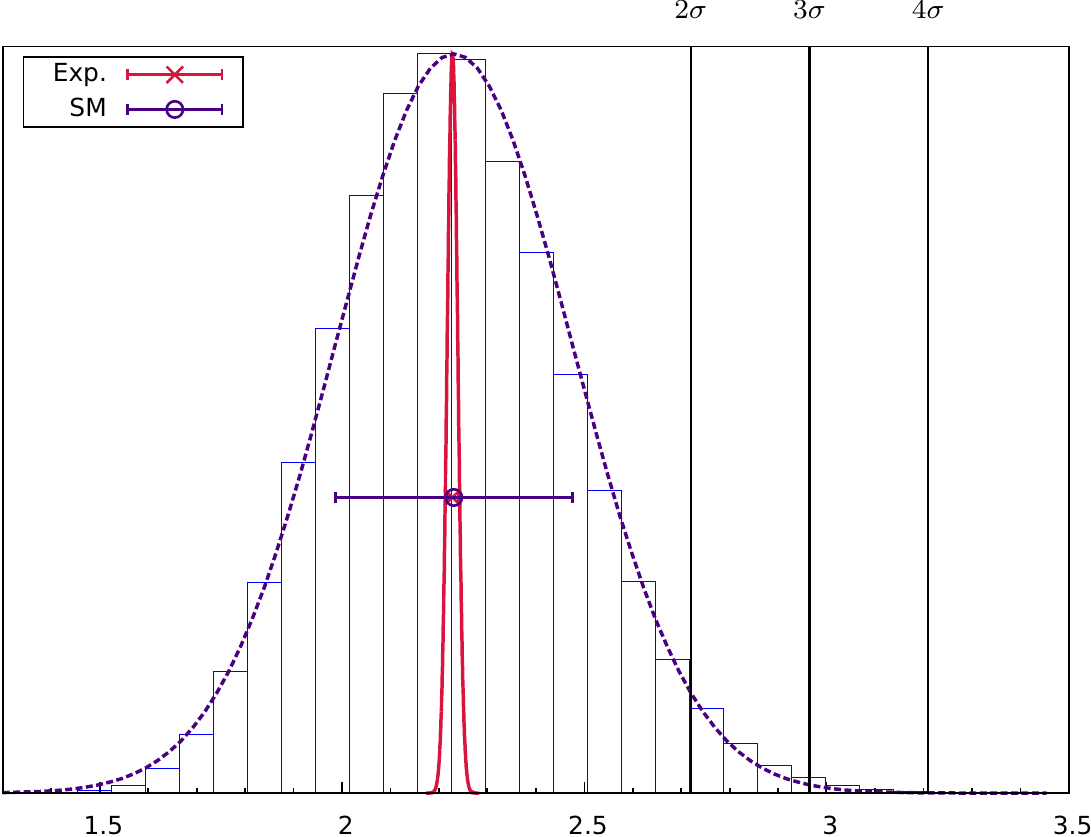}
  }\hspace{0.5cm}
  \subfloat[Excl. $V_{cb}$ + FLAG $\hat{B}_K$\label{sfig:excl.avg.AO}]{%
    \includegraphics[width=0.4\textwidth]{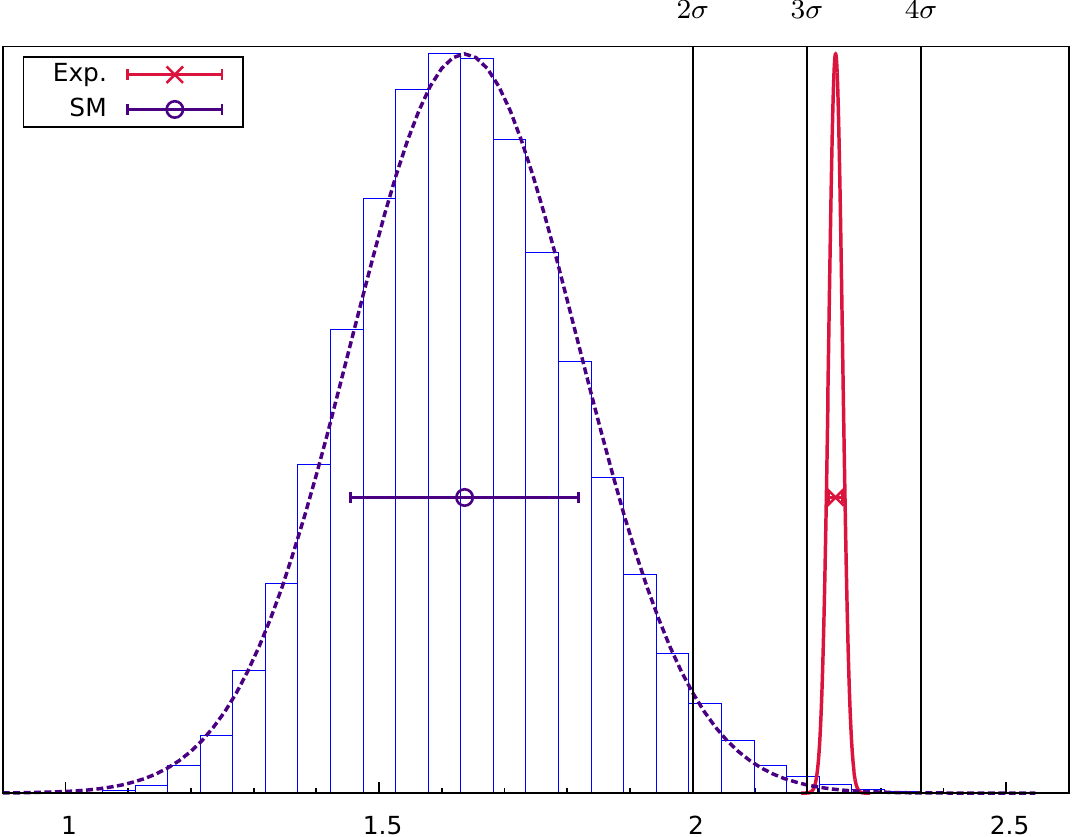}
  }\\
  \subfloat[Incl. $V_{cb}$ + SWME $\hat{B}_K$\label{sfig:incl.SWME.AO}]{%
    \includegraphics[width=0.4\textwidth]{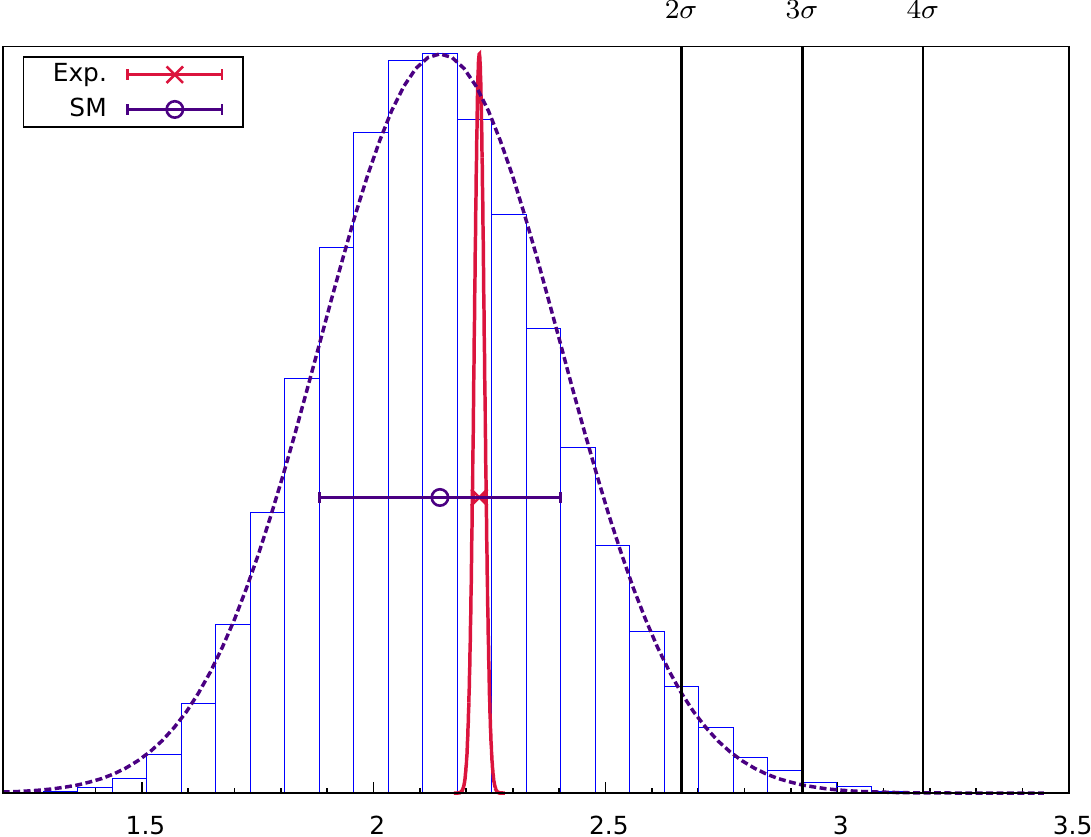}
  }\hspace{0.5cm}
  \subfloat[Excl. $V_{cb}$ + SWME $\hat{B}_K$\label{sfig:excl.SWME.AO}]{%
    \includegraphics[width=0.4\textwidth]{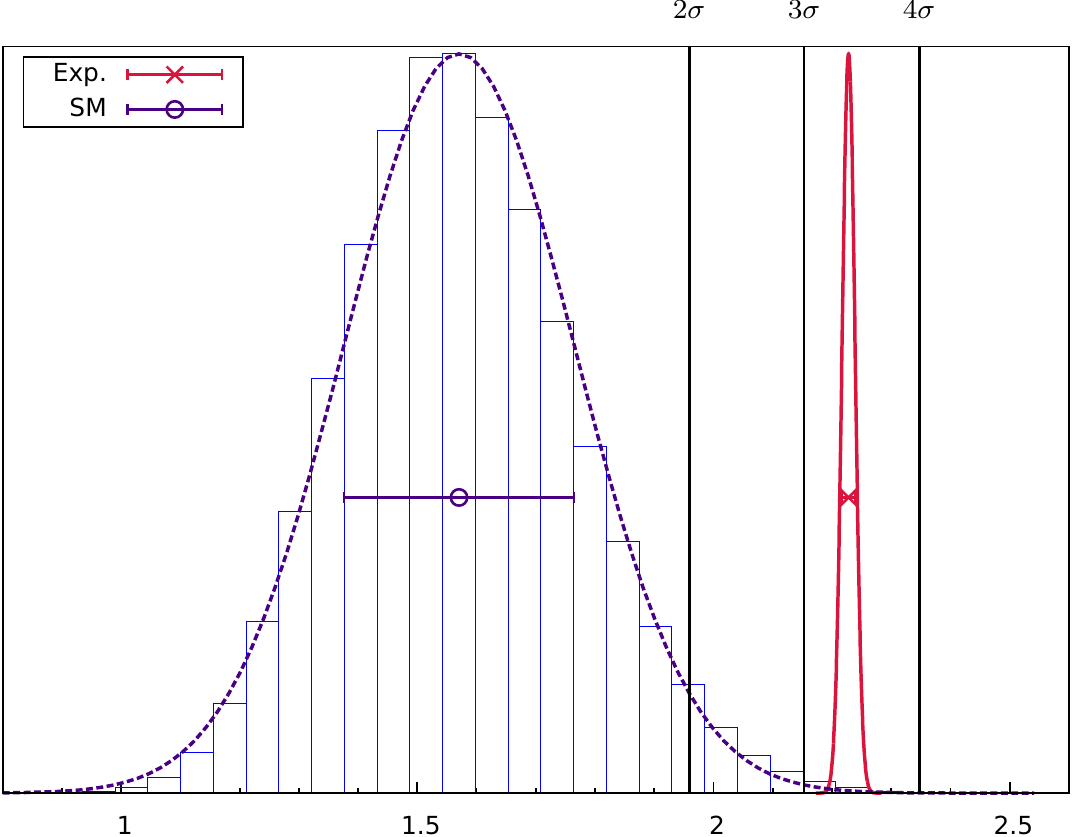}
  }
  \caption{ $\epsK$ with the AOF set of Wolfenstein parameters.  Each
    label shows the combination of $V_{cb}$ and $\hat{B}_K$ inputs.
    The red narrow distribution represents experimental values.  The
    dotted blue wide distribution represents the results of Monte
    Carlo method.  With exclusive $V_{cb}$ we observe a tension
    exceeding $3.1\sigma$, which disappears with inclusive $V_{cb}$.
  }
\label{fig:result.AOF}
\end{figure*}

We compare the SM values of $\epsK$ for our various input choices with
the experimental value in Eq.~\eqref{eq:epsK-exp}.
The Monte Carlo results with the AOF parameter inputs are given in
Fig.~\ref{fig:result.AOF}.
These results are consistent with those obtained using the input
parameters of the CKMfitter and UTfit groups with their implicit
dependence on $\epsK$, $\hat{B}_K$, and $V_{cb}$.
Hence, regardless of the choice of Wolfenstein parameters, the SM is
in good agreement with the experiment, if we use the inclusive
$V_{cb}$.
However, a substantial tension of $3.3(2)\sigma$ between the SM and
the experiment exists with the exclusive $V_{cb}$, AOF inputs, and the
FLAG $\hat{B}_K$.
With input parameters from the global fits (CKMfitter and UTfit), this
tension is relaxed but still exceeds 3.1$\sigma$.
The SM appears to deviate from experiment by $3.1\sigma$ to
$3.4\sigma$; the former comes from taking the CKMfitter and FLAG
$\hat{B}_K$ and the latter from taking the AOF and the SWME
$\hat{B}_K$.
The results are shown in Table \ref{tbl:epsK-result}.
The error budget for the AOF with FLAG $\hat{B}_K$ is given in
Table~\ref{tbl:epsK-budget}.
The uncertainty in the value of $V_{cb}$ dominates the error of the SM
value.
\begin{table}[t!]\footnotesize
  \centering
  \renewcommand{\arraystretch}{1.2}
  \subfloat[$\epsK$\label{tbl:epsK-result}]{
  \begin{tabular}{p{1.2cm}|C{1.5cm}|C{1.5cm}}
  \hline\hline
            & FLAG $\hat{B}_K$ & SWME $\hat{B}_K$ \\
  \hline\hline
  CKMfitter & $1.674(180)$ & $1.607(193)$ \\ \cline{2-3}
  $\lambda,\,\bar{\rho},\,\bar{\eta}$ & $3.1\sigma$ & $3.2\sigma$ \\ \hline
  UTfit & $1.683(178)$ & $1.615(192)$ \\ \cline{2-3}
  $\lambda,\,\bar{\rho},\,\bar{\eta}$ & $3.1\sigma$ & $3.2\sigma$ \\ \hline
  AOF & $1.636(182)$ & $1.570(195)$ \\ \cline{2-3}
  $\lambda,\,\bar{\rho},\,\bar{\eta}$ & $3.3\sigma$ & $3.4\sigma$ \\ \hline
  \hline
  \end{tabular}
  }\hspace{0.5cm}
  \subfloat[Error budget\label{tbl:epsK-budget}]{
  \begin{tabular}{C{1cm}|C{1.2cm}|C{2.5cm}}
    \hline \hline
    source       & error (\%) & memo \\
    \hline
    $V_{cb}$     & 41.3        & FNAL/MILC \\
    $\bar{\eta}$ & 21.7        & AOF \\
    $\eta_3$     & 16.8        & $c-t$ Box \\
    $\eta_1$     &  5.1        & $c-c$ Box \\
    $\bar{\rho}$ &  4.6        & AOF \\
    $m_t$        &  3.4        & \\
    $\xi_0$      &  2.2        & RBC/UKQCD\\
    $\hat{B}_K$  &  1.6        & FLAG \\
    $\vdots$     & $\vdots$    & \\
    \hline \hline
  \end{tabular}
  }
  \caption{(a) $\epsK$ with exclusive $V_{cb}$, and (b) relative error
    budget for the AOF set with FLAG $\hat{B}_K$.}
  \label{tbl:results}
\end{table}


\section{Conclusion}
\label{sec:conclude}
With FLAG average for $\hat{B}_K$ and $V_{cb}$ from the lattice (FNAL/MILC)
form factor for $\bar{B}\to D^*\ell\bar{\nu}$, we find the SM value of
$\epsK$ differs from the experimental value by $3.3(2)\sigma$.
However, with the inclusive $V_{cb}$, we do not observe any tension.
%
%
%
The dominant error in $\epsK$ comes from $V_{cb}$.
New lattice QCD calculations and updated UT analyses are essential.
To contribute to this effort, we plan to calculate the form factors
for $\bar{B}\to D^*\ell\bar{\nu}$ using the Oktay-Kronfeld (OK)
action, which is designed to reduce heavy quark discretization errors
\cite{ OK, MBO:LAT2010, Jang:LAT2014}.

\section{Acknowledgments}

The research of W.~Lee is supported by the Creative Research
Initiatives Program (No.~2014001852) of the NRF grant funded by the
Korean government (MEST).  W.~Lee would like to acknowledge the
support from KISTI supercomputing center through the strategic support
program for the supercomputing application research
[No.~KSC-2013-G2-005].  Computations were carried out on the DAVID GPU
clusters at Seoul National University.  J.A.B. is supported by the
Basic Science Research Program of the National Research Foundation of
Korea (NRF) funded by the Ministry of Education (No.~2014027937).

\bibliography{refs}

\providecommand{\href}[2]{#2}\begingroup\raggedright\begin{thebibliography}{10}

\bibitem{Beringer2012:PhysRevD86.010001}
J.~Beringer {\em et~al.} {\em Phys.Rev.} {\bf D86} (2012) 010001.

\bibitem{Cabibbo1963:PhysRevLett.10.531}
N.~Cabibbo {\em Phys.Rev.Lett.} {\bf 10} (1963) 531--533.

\bibitem{Kobayashi1973:ProgTheorPhys.49.652}
M.~Kobayashi and T.~Maskawa {\em Prog.Theor.Phys.} {\bf 49} (1973) 652--657.

\bibitem{Aoki2013:hep-lat.1310.8555}
S.~Aoki, Y.~Aoki, C.~Bernard, {\em et~al.} {\em Eur.Phys.J.} {\bf C74} (2014),
  no.~9 2890, [\href{http://xxx.lanl.gov/abs/1310.8555}{{\tt 1310.8555}}].

\bibitem{Bae2014:prd.89.074504}
T.~Bae {\em et~al.} {\em Phys.Rev.} {\bf D89} (2014) 074504,
  [\href{http://xxx.lanl.gov/abs/1402.0048}{{\tt 1402.0048}}].

\bibitem{GS}
P.~Gambino and C.~Schwanda {\em Phys.Rev.} {\bf D89} (2014) 014022,
  [\href{http://xxx.lanl.gov/abs/1307.4551}{{\tt 1307.4551}}].

\bibitem{JWL}
J.~A. Bailey, A.~Bazavov, C.~Bernard, {\em et~al.} {\em Phys.Rev.} {\bf D89}
  (2014) 114504, [\href{http://xxx.lanl.gov/abs/1403.0635}{{\tt 1403.0635}}].

\bibitem{Bevan2013:npps241.89}
A.~Bevan, M.~Bona, M.~Ciuchini, {\em et~al.} {\em Nucl.Phys.Proc.Suppl.} {\bf
  241-242} (2013) 89--94.

\bibitem{Durr:2011ap}
S.~Durr, Z.~Fodor, C.~Hoelbling, {\em et~al.} {\em Phys.Lett.} {\bf B705}
  (2011) 477--481, [\href{http://xxx.lanl.gov/abs/1106.3230}{{\tt 1106.3230}}].

\bibitem{Jang2012:PoS.LAT2012.269}
Y.-C. Jang and W.~Lee {\em PoS} {\bf LATTICE2012} (2012) 269,
  [\href{http://xxx.lanl.gov/abs/1211.0792}{{\tt 1211.0792}}].

\bibitem{Wlee}
J.~A. Bailey, Y.-C. Jang, and W.~Lee {\em in preparation}.

\bibitem{Buras1998:hep-ph/9806471}
A.~J. Buras \href{http://xxx.lanl.gov/abs/hep-ph/9806471}{{\tt
  hep-ph/9806471}}.

\bibitem{RBC2014:PRL}
Z.~Bai, N.~Christ, T.~Izubuchi, {\em et~al.} {\em Phys.Rev.Lett.} {\bf 113}
  (2014) 112003, [\href{http://xxx.lanl.gov/abs/1406.0916}{{\tt 1406.0916}}].

\bibitem{Bae2012:PhysRevLett.109.041601}
T.~Bae {\em et~al.} {\em Phys.Rev.Lett.} {\bf 109} (2012) 041601,
  [\href{http://xxx.lanl.gov/abs/1111.5698}{{\tt 1111.5698}}].

\bibitem{rbc-prd-2011-1}
Y.~Aoki, R.~Arthur, T.~Blum, {\em et~al.} {\em Phys.Rev.} {\bf D84} (2011)
  014503, [\href{http://xxx.lanl.gov/abs/1012.4178}{{\tt 1012.4178}}].

\bibitem{alv-prd-2010-1}
C.~Aubin, J.~Laiho, and R.~S. Van~de Water {\em Phys.Rev.} {\bf D81} (2010)
  014507, [\href{http://xxx.lanl.gov/abs/0905.3947}{{\tt 0905.3947}}].

\bibitem{Blum2011:PhysRevLett.108.141601}
T.~Blum, P.~Boyle, N.~Christ, {\em et~al.} {\em Phys.Rev.Lett.} {\bf 108}
  (2012) 141601, [\href{http://xxx.lanl.gov/abs/1111.1699}{{\tt 1111.1699}}].

\bibitem{Brod:2011ty}
J.~Brod and M.~Gorbahn {\em Phys.Rev.Lett.} {\bf 108} (2012) 121801,
  [\href{http://xxx.lanl.gov/abs/1108.2036}{{\tt 1108.2036}}].

\bibitem{Buras:2013raa}
A.~J. Buras and J.~Girrbach {\em Eur.Phys.J.} {\bf C73} (2013) 2560,
  [\href{http://xxx.lanl.gov/abs/1304.6835}{{\tt 1304.6835}}].

\bibitem{Brod2010:prd.82.094026}
J.~Brod and M.~Gorbahn {\em Phys.Rev.} {\bf D82} (2010) 094026,
  [\href{http://xxx.lanl.gov/abs/1007.0684}{{\tt 1007.0684}}].

\bibitem{Alekhin2012:plb.716.214}
S.~Alekhin, A.~Djouadi, and S.~Moch {\em Phys.Lett.} {\bf B716} (2012)
  214--219, [\href{http://xxx.lanl.gov/abs/1207.0980}{{\tt 1207.0980}}].

\bibitem{Buras2008:PhysRevD.78.033005}
A.~J. Buras and D.~Guadagnoli {\em Phys.Rev.} {\bf D78} (2008) 033005,
  [\href{http://xxx.lanl.gov/abs/0805.3887}{{\tt 0805.3887}}].

\bibitem{OK}
M.~B. Oktay and A.~S. Kronfeld {\em Phys.Rev.} {\bf D78} (2008) 014504,
  [\href{http://xxx.lanl.gov/abs/0803.0523}{{\tt 0803.0523}}].

\bibitem{MBO:LAT2010}
C.~Detar, A.~Kronfeld, and M.~Oktay {\em PoS} {\bf LATTICE2010} (2010) 234,
  [\href{http://xxx.lanl.gov/abs/1011.5189}{{\tt 1011.5189}}].

\bibitem{Jang:LAT2014}
J.~A. Bailey, C.~Detar, Y.-C. Jang, A.~Kronfeld, M.~Oktay, and W.~Lee {\em PoS}
  {\bf LATTICE2014} (2014) 097.

\end{thebibliography}\endgroup

\end{document}